\documentclass[twocolumn,prb,showpacs,preprintnumbers,amsmath,amssymb]{revtex4}

\usepackage{graphicx}
\usepackage{dcolumn}
\usepackage{bm}
\usepackage{epsfig}

\begin{document}


\title{Fragility and elasticity - a new perspective}

\author{U. Buchenau}
\email{buchenau-juelich@t-online.de}
\affiliation{%
Institut f\"ur Festk\"orperforschung, Forschungszentrum J\"ulich\\
Postfach 1913, D--52425 J\"ulich, Federal Republic of Germany}%
\date{September 23, 2009}

\begin{abstract}
The fragility (the abnormally strong temperature dependence of the viscosity) of highly viscous liquids is shown to have two sources. The first is the temperature dependence of the barriers between inherent states considered earlier. The second is the recently discovered asymmetry between the actual inherent state and its neighbors. One needs both terms for a quantitative description.
\end{abstract}


\maketitle

Though there is as yet no generally accepted explanation of the flow in highly viscous liquids \cite{heuerr,dyre,ngai}, it seems clear that its description requires the passage of high energy barriers between inherent states, i.e. local structural minima of the potential energy \cite{stillinger}. According to the elastic models \cite{dyre}, the fragility stems from a proportionality of the height $V$ of these barriers to the infinite frequency shear modulus $G$, which in the highly viscous liquid decreases strongly with temperature.

The present paper shows that this is only part of the truth. There is a second source of fragility in the newly discovered \cite{olsen} asymmetry of about $4k_BT$ between the actual inherent state and its neighbors, possibly due \cite{asymm} to the elastic distortion accompanying a structural rearrangement (the "Eshelby backstress" \cite{harmon}). As will be seen, the quantitative explanation of the fragility of six different glass formers requires just this specific explanation of the asymmetry.

The usual measure of the fragility of a glass former is the logarithmic slope of the relaxation time $\tau_\alpha$ of the flow process
\begin{equation}\label{m}
	m=d\log\tau_\alpha/d(T_g/T)|_{T_g},
\end{equation}
where the glass temperature $T_g$ is defined as the temperature with $\tau_\alpha=1000$ s.

It is useful to relate $\tau_\alpha$ to a critical barrier $V_c$ via the Arrhenius relation
\begin{equation}\label{arrh}
	\tau_\alpha=\tau_0\exp(V_c/k_BT),
\end{equation}
where the microscopic attempt frequency is at 10$^{-13}$ s, sixteen decades faster than the flow process at the glass temperature. The {\it fragility index} $I$ is defined \cite{dyre} by the logarithmic derivative $I=-d\ln V_c/d\ln T$, taken at $T_g$. Then
\begin{equation}\label{mI}
	m=16(I+1),
\end{equation}
where the factor reflects the sixteen decades between microscopic and macroscopic time scales. $I$ is a better measure of the fragility than $m$, because it does not contain the trivial temperature dependence of any thermally activated process.

The elastic models \cite{dyre} postulate a proportionality between the flow barrier $V_c$ and the infinite frequency shear modulus $G$. One can again define a dimensionless measure $\Gamma$ for the temperature dependence of $G$ in terms of the logarithmic derivative $\Gamma=-d\ln G/d\ln T$ at $T_g$. Then the elastic models \cite{dyre} postulate $I=\Gamma$. 

In order to check this relation, one needs measurements of both quantities. The flow relaxation time $\tau_\alpha$ is relatively easy to measure, but the determination of the high frequency shear modulus is by no means trivial. It requires the measurement of the density $\rho$ and the high frequency transverse sound velocity $v_t$. Consequently, the logarithmic derivative $\Gamma$ is a sum of two terms, a larger one from the sound velocity and a smaller one from the density.

In a liquid, well-defined transverse sound waves do only exist at frequencies which are markedly higher than the inverse $1/\tau_\alpha$ of the flow relaxation time. With increasing temperature, $\tau_\alpha$ gets very rapidly shorter. Therefore the measurement of the transverse sound velocity by Brillouin scattering is limited to a small temperature region above $T_g$. This, together with the poor visibility of the transverse Brillouin line, leads to a large error bar in the determination of $\Gamma$, usually about 20 \%.

\begin{table}[htbp]
	\centering
		\begin{tabular}{|c|c|c|c|c|c|c|c|c|c|}
\hline
subst.   & $T_g$ & $G$    &   $m$   &  $I$ & $\Gamma$   & $f_cV_c$ & $\alpha_VT_g$ &   $I_1$ & $I_2$ \\
\hline   
         &   $K$  & $GPa$ &         &      &            &          &               &         &       \\
\hline                                                                                           
silica   & 1449   &  31   &  28     & 0.5  & 0.07       & 3.61     &   0           &   0.37  &  0.3  \\
Vit-4    &  627   &  34   &  30     & 0.88 & 0.56       &          &               &         &       \\
glycerol &  187   &  4.5  &  53     & 2.31 & 1.0        & 1.59     &  0.12         &   2.2   &  1.2  \\
PB20     &  173   &  1.8  &  84     & 4.25 & 2.0        & 1.05     &  0.12         &   4.8   &  2.9  \\
CKN      &  343   &  4.9  &  93     & 4.81 & 2.6        &          &               &         &       \\
BPA-PC   &  418   &  0.8  & 132     & 7.25 & 4.4        & 1.51     &  0.23         &   7.8   &  3.4  \\
PS       &  375   &  1.0  & 138     & 7.63 & 4.0        & 0.82     &  0.21         &   9.8   &  5.8  \\
PMMA     &  379   &  1.9  & 145     & 8.06 & 2.1        & 0.30     &  0.23         &  11.7   &  9.6  \\
\hline		
		\end{tabular}
	\caption{Measured and calculated fragilities for eight glass formers. Vit-4 is a bulk metallic glass, PB20 is a 20:80 mixture of 1,2-polybutadiene and 1,4-polybutadiene, CKN is K$_3$Ca$_2$(NO$_3$)$_7$, BPA-PC is bisphenol-A-polycarbonate, PS is polystyrene and PMMA is polymethylmethacrylate. References see ref. \cite{ta}.}
	\label{tab:Comp}
\end{table}

In spite of these difficulties, six apparently reliable measurements of $\Gamma$ by Brillouin scattering exist in the literature. The six substances are listed in Table I, together with the bulk metallic glass Vit-4 and polystyrene, where $\Gamma$ was determined from ultrasonic measurements.

Table I compares fragility indices $I$ calculated from the temperature dependence of $\tau_\alpha$ with $\Gamma$. One finds that $I$ is always larger than $\Gamma$, in several cases clearly beyond the estimated error bar of 20 \%. This is very surprising, because one would have expected the opposite result. The barriers of the energy landscape in molecular glass formers are frequently intramolecular barriers \cite{paluch}, much less temperature dependent than the van-der-Waals dominated shear modulus. The same holds for polymers, where the torsional barriers \cite{paul,bernabei} are practically temperature-independent. Thus one would expect $I<\Gamma$, but one finds $\Gamma<I$. Though there is a clear tendency of a fragility increase with increasing $\Gamma$, the postulate $V_c\sim G$ is obviously not sufficient to explain the full observed fragility.

A second possible source of fragility is the recently discovered asymmetry between the actual inherent state and its neighbors \cite{olsen}. The strength of an asymmetric relaxation increases with increasing temperature, because the thermal population of the upper level increases.

In order to quantify this influence within the asymmetry model \cite{asymm}, an extension of the coupling model \cite{ngai}, consider its basic definition of the characteristic multi-minimum parameter $f_N$
\begin{equation}\label{fn}
    f_N=\frac{c_N}{N^3}\left(\frac{k_BT}{Gv}\right)^3,
\end{equation}
where $T$ is the temperature and $v$ is the atomic or molecular volume. $c_N$ is a temperature-independent measure of the density of stable states for $N$ atoms or molecules in distortion space, assumed to be constant. $N$ must be large enough to meet the condition $f_N=1$ for the breakdown of the shear modulus at $T_g$. The barrier $V_c$ is the lowest barrier for all possible $N$ to reach $f_N=1$. In terms of the definitions of the coupling model \cite{ngai}, the jumps into neighboring inherent states with barriers below $V_c$ are the primitive relaxations. The asymmetry model \cite{asymm} postulates that their elastic interaction brings the shear modulus down to zero.

If this is indeed so, the contribution to the fragility depends crucially on the barrier density of primitive relaxations at $V_c$. In the model, the primitive relaxation density is characterized by a barrier density function $f_0(V)$ and $V_c$ is given by the 1/3-rule
\begin{equation}\label{vc}
	\int_0^{V_c} f_0(V)=\frac{1}{3}.
\end{equation}

According to its definition in terms of $f_N$, $f_0(V)$ increases with increasing temperature proportional to $(T/Gv)^3$. This shifts $V_c$ downwards and so provides an additional fragility index $I_2$
\begin{equation}\label{I2}
	I_2=\frac{1+\Gamma-\alpha_VT_g}{f_cV_c}.
\end{equation}
Here $\alpha_V$ is the volume expansion coefficient above the glass temperature $T_g$ and $f_c=f_0(V_c)$.

\begin{figure}[b]
\hspace{-0cm} \vspace{0cm} \epsfig{file=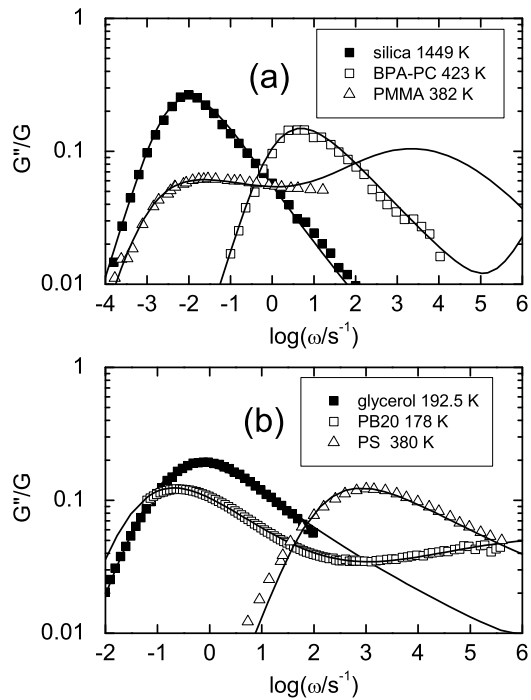,width=7cm,angle=0} \vspace{0cm}\caption{Fits of dynamical shear data in terms of the asymmetry model \cite{asymm} (continuous lines) for (a) silica, BPA-PC and PMMA (b) glycerol, PB20 and polystyrene. Note that a high $f_c$ means a high value of $G''/G$ at the $\alpha$-peak. Position and width of the secondary peaks of BPA-PC and PMMA were taken from mechanical relaxation data in the glass phase.}
\end{figure}

In order to determine $f_c$, one needs an asymmetry model fit of dynamical shear data for the given substance, preferably close to the glass temperature, but still in the equilibrium liquid. For this purpose, one can follow the recipe given in the asymmetry paper \cite{asymm}, describing the dynamical shear data with the three parameters $G$, $V_c$ and $\gamma$ ($\gamma$ describes the exponential rise $f_0(V)\sim \exp(\gamma V/k_BT)$ of $f_0(V)$ at $V_c$).  For $G$, one takes the measured high-frequency value from Table I. If necessary, one can add a gaussian to describe an eventual Johari-Goldstein peak \cite{johari}, which requires three more parameters, height, position and width. In this way, $f_cV_c$ and consequently $I_2$ were obtained for the six glass formers in Table I for which dynamical shear data close to $T_g$ exist in the literature \cite{ta}. Fig. 1 (a) and (b) show the fits of the dynamical shear data.

Table I compares $I_2$ and the supposedly full fragility $I_1=\Gamma+I_2$ to the measured value $I$. It is immediately clear that one needs $I_2$ to understand the full measured fragility. But sometimes $I_1$ is a bit too large, particularly in polystyrene and in PMMA. This is not unexpected, because the elastic model \cite{dyre} expectation $V\sim G$ holds only in substances where the contribution of intramolecular barriers is negligible \cite{paluch}. In polymers, the torsional barriers do play a role \cite{paul,bernabei} and do not share the temperature dependence of $G$. In PMMA, where the relaxation is dominated by a side group relaxation, one should rather compare $I$ with $I_2$. In fact, in this case $I_2$ agrees with $I$ within the error bars of the three measurements involved \cite{ta}.

From eq. (\ref{I2}), one sees that the second contribution to the fragility is weak whenever $f_c$ is high. This is often the case in molecular glass formers (glycerol is an exception). For instance, in triphenylethylene \cite{asymm} $f_c=5.31$, markedly higher than the one in silica. This explains why a reasonable agreement with the elastic model postulate $V_c\sim G$ alone has been found in molecular glasses \cite{shoving,nelson}. But even in these two papers, there are several cases which clearly have a higher fragility.

The elastic model postulate is the most obvious connection between fragility and fast vibrations, but it is by no means the only proposition in this direction. Other proposals relate the fragility to the nonergodicity factor measured in high-resolution x-ray scattering \cite{scoscience} or to the Poisson ratio \cite{sokolov}. The relation to the nonergodicity factor is understandable, because a low nonergodicity factor means a low level of density fluctuations, which in turn means that one is close to the ideal glass of the Kauzmann paradoxon and expects a high thermodynamic fragility \cite{ra}. Very recently \cite{scorome}, it has been pointed out that there are exceptions from the nonergodicity rule due to a strong influence of secondary relaxations \cite{cangialosi}, a reasoning which is parallel to the one in the present work, a second fragility influence which requires not only a knowledge of the fast motion, but also of the relaxations themselves.

To summarize: A full quantitative understanding of the fragility requires the consideration of both the temperature dependence of the barriers and the influence of the recently discovered \cite{olsen} asymmetry of an actually occupied inherent state with respect to its neighbors. The latter plays a minor role if the density of relaxations at the critical flow barrier is high, but can become dominant if it happens to be low. The quantitative agreement supports the validity of the explanation of the asymmetry in terms of an elastic distortion \cite{asymm}, an Eshelby backstress \cite{harmon} which tends to stabilize the occupied inherent states.

\end{document}